\documentstyle[amsmath,amssymb,graphicx]{article}

\def\be{\begin{eqnarray}}
\def\ee{\end{eqnarray}}
\def\nn{\nonumber}

\def\p{\partial}

\def\l[{\phantom.[}

\def\BS{\bar{\cal S}}

\def\embed{\longrightarrow}

\hoffset= - 1.0in         

\begin{document}

\hfill ITEP/TH-01/17

\hfill IITP/TH-01/17

\bigskip

\centerline{\Large{
On moduli space of symmetric orthogonal matrices and
}}
\centerline{\Large{
exclusive Racah matrix $\bar S$ for representation $R=[3,1]$
with multiplicities
}}

\bigskip

\centerline{\bf  A.Morozov }

\bigskip

{\footnotesize
\centerline{{\it
ITEP, Moscow 117218, Russia}}

\centerline{{\it
Institute for Information Transmission Problems, Moscow 127994, Russia
}}

\centerline{{\it
National Research Nuclear University MEPhI, Moscow 115409, Russia
}}
}

\bigskip

\centerline{ABSTRACT}

 \bigskip

{\footnotesize
Racah matrices and higher $j$-symbols are used in description of
braiding properties of conformal blocks and in construction of
knot polynomials.
However, in complicated cases the logic is actually inverted:
they are much better {\it deduced} from these
applications than from the basic representation theory.
Following the recent proposal of
\cite{de31}
we  obtain  the exclusive Racah matrix $\bar S$
for the currently-front-line case of representation $R=[3,1]$
with non-trivial multiplicities, where it is actually
operator-valued, i.e. depends on the choice of basises
in the intertwiner spaces.
Effective field theory for arborescent knots in this case possesses
gauge invariance, which is not yet properly described and understood.
Because of this lack of knowledge a big part (about a half)
of $\bar S$ needs to be reconstructed from orthogonality conditions.
Therefore we discuss the abundance of {\it symmetric} orthogonal matrices,
to which $\bar S$ belong, and explain that dimension of their moduli space is
also about a half of that for  the ordinary orthogonal matrices.
Thus the knowledge approximately matches the freedom  and this explains
why the method can work -- with some limited addition of educated guesses.
A similar calculation for $R=[r,1]$ for $r>3$ should also be doable.
}

\bigskip

\section{Introduction}

Racah matrices (also known as $6j$-symbols) are a traditional topic
in theoretical and mathematical physics,
with a special chapter dedicated to them already in \cite{LL3}.
Despite a long history of research and with all available computer power,
actual computation of these quantities
remains among the most difficult problems, and until very recently
all the non-trivial and interesting examples were  out of reach.
In modern theory $j$-symbols appear in two intimately related stories:
modular transformations of conformal blocks and
evaluation of physical observables in Chern-Simons theory \cite{Wit}
(known as Wilson loop averages or knot polynomials \cite{knotpols}).
As often happens, these {\it applications} of Racah theory actually
provide the most efficient way to {\it calculate} them.
The present paper is a one more illustration of this inverse feedback
from physical would-be-applications to basic mathematics:
It reports a new breakthrough in Racah calculus --
to  series $[r,1]$ of representations with multiplicities,
and Racah matrices are extracted from a new deep knowledge about knot
polynomials -- the structure of their differential expansion.

We do not go into details about knots, referring the interested reader
to \cite{de31} and references therein.
Instead we concentrate on the complementary part of the story,
coming from the fact that one of the relevant Racah matrices, called $\bar S$,
 is actually a little peculiar: it is orthogonal, as any properly normalized
$6j$-symbol (when real-valued, in general it is unitary),
but at the same time it is symmetric.
Intersection of these two requirements actually restricts a matrix a lot --
and this allows to reconstruct it from a fragment.
A fragment is exactly what is currently known about $\bar S$ in
representations $R=[3,1]$ from knot theory -- and it is of approximately the
right size which is necessary for the reconstruction.
Matching is not exact and, more important, not quite under control,
because it is not clear how to separate the independent orthogonality
constraints -- but it is at least a motivation for a try.
In fact, one can add intuition of another kind: Racah matrices
usually depend on the quantum group/knot theory parameters $q$ and $A$
in a relatively nice way: many of matrix elements factorize into
products/ratios of "differentials" (actually, of quantum numbers),
and those which do not, deviate from factorized form only "moderately".
It is highly non-trivial to get such nearly-factorized quantities
satisfying non-linear orthogonality relations -- and this imposes
additional strong constraints, which, however, are still very difficult
to formalize.
In this paper we report the result of a tedious analysis, leading to
a very plausible answer for $\bar S$ in representation $[3,1]$.
It generalizes the celebrated result of \cite{GJ} for $R=[2,1]$,
which was obtained from the first principles in a sophisticated 70-page paper,
but became nearly trivial in the approach of \cite{de31}.
The new formula is tested by providing  polynomial expressions
for HOMFLY-PT Wilson-loop averages for numerous knots,
some of which (for 2-strand knots) are actually known from other sources.
In principle, one can now build a second exclusive matrix $S$
and apply the machinery of \cite{arbor} to do
calculations for all arborescent knots \cite{arbotexts}.
This is an important task, because the arborescent calculus of \cite{arbor} is based
on a very interesting effective field theory, which possesses a peculiar
gauge invariance, associated with {\it multiplicities} in representation theory,
and which is not yet satisfactorily formulated.
One can expect that multiplicity problem does not arise to its full size
for representations smaller than $R=[4,2]$, because gauge invariance for
them is actually partly broken by diagonal matrices $T$ and $\bar T$ --
this appeared to be the case for $R=[2,1]$, but remains to be tested for $R=[3,1]$.
This test is made possible by the result/conjecture of the present paper,
but it is left for the future work.

In this paper we concentrate on the problem of its own:
evaluation of $\bar S_{[3,1]}$ in a particular basis.
We begin from reminding the notion of Racah matrices in sec.2,
then discuss the moduli space of symmetric orthogonal matrices in sec.3.
After that in sec.4 we briefly comment on the calculation, suggested
in \cite{de31}, which includes the clever choice of a basis --
expressed in the form of a special ansatz for the shape of $\bar S$.
The complement of the piece $\bar{\cal S}\subset \bar S$,
which was earlier found in \cite{de31}, is provided in explicit form in
the Appendix, the full matrix is available -- together with all other
currently known examples -- at the site \cite{knotebook}.
A very brief description of immediate knot theory applications is
provided in sec.5.

\section{The options for Racah calculus}

\subsection{Racah matrices}

The product of $m$ irreducible representations $R_i$ of a Lie algebra
${\cal G}$ (classical or quantum) can be decomposed into a linear
combination of irreps:
\be
\otimes_{i=1}^m R_i = \oplus_Q \ W_Q^{R_1,\ldots, R_m}\otimes Q
\ee
If representation $Q$ appears at the r.h.s. with non-trivial multiplicity,
then there is a space $W_Q$ of intertwiners,
which is representation of the symmetry group $S_m$.
Racah matrix $U$ describes a liner map between the spaces $W^{(3)}$
and it intertwines
$(R_1\otimes R_2)\otimes R_3 \longrightarrow Q$
and $R_1\otimes (R_2\otimes R_3)\longrightarrow Q$:

\begin{picture}(300,125)(-100,-50)
\put(0,0){\line(0,-1){30}}
\put(0,0){\line(-1,1){50}}
\put(0,0){\line(1,1){50}}
\put(-25,25){\line(1,1){25}}
\put(-55,55){\mbox{$R_1$}}
\put(-3,55){\mbox{$R_2$}}
\put(51,55){\mbox{$R_3$}}
\put(5,-35){\mbox{$Q$}}
\put(-10,13){\mbox{$Y$}}
\put(-29,15){\mbox{$a$}}
\put(-11,-5){\mbox{$b$}}
\put(250,0){
\put(0,0){\line(0,-1){30}}
\put(0,0){\line(-1,1){50}}
\put(0,0){\line(1,1){50}}
\put(25,25){\line(-1,1){25}}
\put(-55,55){\mbox{$R_1$}}
\put(-3,55){\mbox{$R_2$}}
\put(51,55){\mbox{$R_3$}}
\put(5,-35){\mbox{$Q$}}
\put(3,13){\mbox{$Z$}}
\put(25,15){\mbox{$c$}}
\put(6,-5){\mbox{$d$}}
}
\put(70,10){\mbox{$= \ \sum_{Z,c,d} \
U\left[\begin{array}{cc}R_2& R_3\\R_1 & Q   \end{array}\right]_{Yab,Zcd} $}}
\end{picture}

\noindent
The labeling of $U$ looks natural in another pictorial representation,
familiar from the study of dualities:

\begin{picture}(300,120)(-66,-60)
\put(0,0){\line(1,0){60}}
\put(0,0){\line(-1,1){25}}
\put(0,0){\line(-1,-1){25}}
\put(60,0){\line(1,1){25}}
\put(60,0){\line(1,-1){25}}
\put(-32,29){\mbox{$R_2$}}
\put(85,29){\mbox{$R_3$}}
\put(-32,-34){\mbox{$R_1$}}
\put(85,-34){\mbox{$Q$}}
\put(28,5){\mbox{$Y$}}
\put(3,5){\mbox{$a$}}
\put(50,5){\mbox{$b$}}
\put(105,-3){\mbox{$= \ \ \sum_{Z,c,d} \ \
U\left[\begin{array}{cc}R_2& R_3\\R_1 & Q   \end{array}\right]_{Yab,Zcd} $}}
\put(280,-15){
\put(0,0){\line(0,1){30}}
\put(0,30){\line(-1,1){25}}
\put(0,30){\line(1,1){25}}
\put(0,0){\line(-1,-1){25}}
\put(0,0){\line(1,-1){25}}
\put(-34,59){\mbox{$R_2$}}
\put(30,59){\mbox{$R_3$}}
\put(-34,-34){\mbox{$R_1$}}
\put(30,-34){\mbox{$Q$}}
\put(3,10){\mbox{$Z$}}
\put(-8,1){\mbox{$c$}}
\put(-8,23){\mbox{$d$}}
}
\end{picture}

\noindent
A basis in $W_Q^{(3)}$ is naturally labeled by the "intermediate" representations
$Y$ or $Z$ in $R_1\otimes R_2$, thus $U$ is a matrix with the indices $Y$ and $Z$.
If not only $W^{(3)}$ at the level of triple products, but also $W^{(2)}$
for ordinary products is non-trivial,
then matrix elements $U_{YZ}$ are actually linear operators, acting from
$W^{R_1,R_2}_Y$ to $W^{R_2,R_3}_{Z}$ and there are additional {\it pairs}
of indices $(ab)$ and $(cd)$ with $a,b=1,\ldots, {\rm dim}\ W^{R_1,R_2}_Y$
and $c,d=1,\ldots,{\rm dim}\ W^{R_2,R_3}_{Z}$ .
Such operator-valued matrices still do not have a commonly-accepted
description, and this -- along with extreme calculational difficulties --
explains the lack of results in the literature.

Various $j$-symbols can be considered as the {\it mixing matrices} \cite{mmmkn1}
between the ${\cal R}$-matrices, which are the generators of the
braid group ${\cal B}_m$, e.g.
$${\cal R}^{(2)} = U{\cal R}^{(1)} U^\dagger$$
Yang-Baxter (braid group) relation
$${\cal R}^{(1)}{\cal R}^{(2)}{\cal R}^{(1)}
= {\cal R}^{(2)}{\cal R}^{(1)}{\cal R}^{(2)}$$
then implies expression for $U$ through ${\cal R}$, like the
{\it eigenvalue hypothesis} \cite{evhyp,TW}.

\subsection{The highest weight method \cite{hw}}

This is the simplest straightforward approach to evaluation of $j$-symbols.
One just explicitly describes highest weights $h_Q$ within the
Verma modules $(R_1\otimes R_2)\otimes R_3$ and $R_1\otimes (R_2 \otimes R_3)$
and then compares.

For example, one can describe the fundamental representation $[1]=V_0$
of $SL_\infty$ by the highest weight $|0>$ and the action of simple roots
$$|k-1\rangle\ \stackrel{T^+_k}{\longrightarrow}\ \delta_{i,k}\cdot  |k\rangle$$
(for the sake of  brevity we omit the group-dependent coefficients,
which can be easily restored).
Then
$$ [1]\otimes [1] = V_{00} \oplus V_{[10]}$$
is a combination of two representations with highest weights
$|0\rangle\otimes |0\rangle$ and
$|1\rangle\otimes |0\rangle - q|0\rangle\otimes |1\rangle$.
At the next stage schematically
\be
\Big([1]\otimes [1]\Big)\otimes [1] = V_{00}\otimes V_0 \oplus V_{[10]}\otimes V_0
= V_{000} \oplus \underline{V_{(10)0-[2]\cdot 001}\oplus V_{[10]0}}\oplus V_{(210)}
\ee
and
\be
[1]\otimes \Big([1]\otimes [1]\Big)  = V_0\otimes V_{00}\oplus \otimes V_0 \otimes V_{[10]}
= V_{000} \oplus \underline{V_{0(10)-[2]\cdot 100}\oplus V_{0[10]}}\oplus V_{(210)}
\ee
where $(\ )$ and $[\ ]$ denote $q$-symmetrization and $q$-antisymmetrization.
Clearly the underlined highest weights in the two cases are different
and Racah matrix relates them (properly normalized)
\be
\left(\begin{array}{c}
\frac{ |100>+\frac{1}{q}|010> - q[2]\cdot |001>}{\sqrt{[2][3]}} \\ \\
\frac{|100>-q|010>}{\sqrt{[2]}}
\end{array}\right) =
\left(\begin{array}{cc}
\frac{1}{[2]} & \frac{\sqrt{[3]}}{[2]} \\ \\
\frac{\sqrt{[3]}}{[2]} & -\frac{1}{[2]}
\end{array}\right)
\left(\begin{array}{c}
\frac{   [2]\cdot |100>-q^2|010>-q|001>}{\sqrt{[2][3]}} \\ \\
\frac{|010>-q|001>}{\sqrt{[2]}}
\end{array}\right)
\ee

Unfortunately, complexity of calculations rapidly grows with the
size of representations.
Situation can be  improved by more advanced description of
highest weights, say, by ($q$-deformed) Vandermonde products \cite{hw}
and Gelfand-Zeitlin labeling \cite{ShSl}, -- but only partly.
Currently, the top achievements on this way is evaluation of
inclusive Racah matrices for representations up to $R=[4,2]$.

\subsection{Conformal block monodromies \cite{Gal}
and exclusive matrices $\bar S$, $S$}

A potentially competitive method uses advances in
the theory of conformal blocks.
Since they can be represented by (appropriately defined)
Dotsenko-Fateev integrals/sums \cite{Sh} and thus belong
to a class of $q$-hypergeometric functions,
their modular properties, which are controlled by the $j$-symbols,
 should be comprehensible.
Advantage of this approach is a relatively simple dependence of vertex operators
on representation, what gives a chance to get formulas for entire
classes of representations at once.
For an example of this kind for $q=1$ (i.e. for the
central charge $c=\infty$, when multiple integrals
are not always needed \cite{Zam,MirMorcinf}) see \cite{Shcinf}.


The simplest of all are the 4-point conformal blocks with
two vertices in representation $R$ and two in the conjugate representation
$\bar R$. The corresponding $6j$-symbols are now called
{\it exclusive} Racah matrices  $\bar S$ and $S$:
\be
\bar S_R : \ \ \ \ \Big((R\otimes \bar R)\otimes R\ \longrightarrow \ R\Big)
\ \ \ \longrightarrow \ \ \
\Big(R\otimes (\bar R\otimes  R)\ \longrightarrow \ R\Big)
\ee
and
\be
S_R : \ \ \ \ \Big((R\otimes R)\otimes \bar R\ \longrightarrow \ R\Big)
\ \ \ \longrightarrow \ \ \
\Big(R\otimes ( R\otimes  \bar R)\ \longrightarrow \ R\Big)
\ee
They are difficult to calculate by the highest weight method,
because highest weights of the conjugate representations depend strongly
on the choice of the group $SL_N$ -- therefore one needs to calculate for
different values of $N$ and then analytically continue.

Instead these matrices can be looked for by the {\it evolution method}
in knot theory \cite{evo,mmms3}.
This paper describes a new achievement of this approach --
for representations $R=[r,1]$ where multiplicities begin to matter.
We immediately reproduce in this way the difficult result of \cite{GJ} for $R=[2,1]$
and conjecture the answer for  $R=[3,1]$.
This adds to the previously known cases of arbitrary symmetric
representation $R=[r]$ in \cite{IMMMfe,symreps} and rectangular representations
$R=[r^s]$ in \cite{mmms3,db} (in the latter case actually tabulated are
Racah matrices for the two-line $R=[rr]$ with $r\leq 5$, see \cite{knotebook}).
Formulas for transposed representations (say, antisymmetric or two-column)
are obtained by the change $q\longrightarrow -q^{-1}$ \cite{DMMSS}.

\section{The abundance of matrices $\bar S$ and $S$}

\subsection{Yang-Baxter relation}

 Because of the Yang-Baxter relation the matrices $S$ and $\bar S$
 are not independent.
 If we denote the diagonalized ${\cal R}$ matrices in the channels
 $R\otimes R$ and $R\otimes\bar R$ by $T$ and $\bar T$ respectively,
 then
 \be
 S^\dagger T S = \bar T^{-1}\bar S\bar T^{-1}
 \label{SthrbS}
 \ee
 Moreover, by its definition $\bar S$ is a {\it symmetric} orthogonal matrix,
 thus (\ref{SthrbS}) defines $S$ as the diagonalizing matrix of
 symmetric (but no longer orthogonal) $T^{-1}\bar S\bar T^{-1}$,
 i.e. $S$ defines $\bar S$ and vice versa
 -- for given diagonal $T$ and $\bar T$ with no degenerate eigenvalues.
 If $T=\bar T$ and $S=\bar S$, then (\ref{SthrbS}) becomes a non-trivial
 quadratic relation, presumably leading to the {\it eigenvalue hypothesis}
 \cite{evhyp,TW}.
 Degeneration of eigenvalues of $T$ and $\bar T$ signals about
 non-trivial {\it multiplicities}, though the situation is somewhat
 more involved: there can be ``{\it accidental}'' degeneracies,
 unrelated to multiplicities (at least in an obvious way) and
 conversely, there can be multiplicities, but no degeneracies
 (eigenvalues can still differ by a sign) -- both phenomena will
 show up in the discussion or representations $R=[r,1]$ in this paper.

 \subsection{Moduli space of symmetric orthogonal matrices}

 For ordinary orthogonal matrices of the size ${\cal N}\times{\cal N}$
 one usually imposes $\frac{{\cal N}({\cal N}+1)}{2}$
 orthonormality constraints on ${\cal N}^2$ elements,
 and if the constraints are all independent this leaves
 ${\cal N}^2 - \frac{{\cal N}({\cal N}+1)}{2} = \frac{{\cal N}({\cal N}-1)}{2}$
 free parameters.
 The simplest way to justify this is just to note that for any
 antisymmetric matrix  $\exp(antisymmetric) = orthogonal$.

 However, such exponentiation will never produce a {\it symmetric} matrix
 (with the only exception of unity), i.e. symmetric orthogonal matrices
 do not possess exponential realization.
 Already for ${\cal N}=2$ they have a form $\sigma_3 \cdot e^{i\alpha\sigma_2}$
 rather than $e^{i\alpha\sigma_2}$ -- and this example is enough to demonstrate
 that now of $\frac{{\cal N}({\cal N}+1)}{2}$ orthogonality constraints
 on $\frac{{\cal N}({\cal N}+1)}{2}$ elements are not always independent.
 If they were, there would be no free parameters (moduli) at all,
 but in fact the set of symmetric orthogonal matrices,
 to which $\bar S$ belongs, is just small.

 \subsection{Eigenvalues and signature of $\bar S$}

Racah matrices, needed in knot theory, are functions of parameters $q$ and $A=q^N$,
which can be arbitrary complex numbers.
However, since the final quantities made out of them are Laurent polynomials,
one can easily continue from the domain where the matrix in a particular
representation $R$ is real-valued (for this one should just keep $A$ and $q$
real and $|A|>|q|^{\pm |R|}$).
Real valued {\it symmetric} matrix $\bar S$ can be diagonalized by conjugation
with orthogonal matrix and has real eigenvalues.
Since $\bar S$ is at the same time orthogonal, these eigenvalues can be only $\pm 1$.
Naturally the spaces of such matrices are classified by their signatures --
the difference between the numbers of eigenvalues $+1$ and $-1$,
and dimension of the moduli space of symmetric orthogonal matrices
depends on the signature.
If all eigenvalues are the same, there are no moduli:
orthogonal conjugate of unit matrix is unit matrix itself.

Since eigenvalues do not depend on $q$, they can be evaluated at $q=1$,
when diagonal $\bar T$ is also made from $\pm 1$ and the eigenvalues of
$\bar S$ merge with those of $\bar T^{-1}\bar S \bar T^{-1}$,
which, according to (\ref{SthrbS}), are just the elements of diagonal $T$.
This means that for every Racah matrix $\bar S$ we actually know its
signature -- it coincides with the signature of $T$.
For example, for all symmetric representations $R=[r]$ the eigenvalues
of $\bar S$ are just an alternating sequence $+1,-1,+1,-1,\ldots$
thus signature is $0$ and $1$ for even and odd ${\cal N}=r+1$ respectively,
while signature $-1$ does not appear.

 \subsection{The elementary cases of ${\cal N}=3$ and ${\cal N}=2$}

 For example, for ${\cal N}=3$ the condition
 \be
 \left(\begin{array}{ccc}
 d_1& a&b \\ a & d_2 & c \\ b&c& d_3
 \end{array}\right)^2 = I
 \ee
 implies
 \be
 a^2 = (d_1+d_3)(d_2+d_3) \nn \\
 b^2 = (d_1+d_2)(d_2+d_3) \nn \\
 c^2 = (d_1+d_2)(d_1+d_3)
 \ee
 and
 \be
 (d_1+d_2+d_3)^2=1
 \ee
 what leaves a 2-parametric set, which for ${\cal N}=2$
 ($c=0$, $d_3=1$) reduces to a 1-parametric
 \be
 \left(\begin{array}{cc}  \cos\theta & \sin\theta \\ \sin\theta & -\cos\theta
 \end{array}\right)
 \label{gbs2}
 \ee
 (note that this is a rotation, complemented by a reflection, and determinant of
 the matrix is $-1$ rather than $1$).

 One can instead express the entries of the symmetric orthogonal matrix
 through those in the first line, satisfying $a^2+b^2+d_1^2=1$:
 \be
 c = \pm\frac{ab}{1\pm d_1} = \pm \frac{ab(1 \mp d_1 )}{1-d_1^2}\nn \\
 d_2 = -d_1 \mp \frac{b^2}{1\pm d_1} = \mp 1 \pm \frac{a^2}{1\pm d_1}
 = - \frac{a^2d_1\pm b^2}{1-d_1^2} \\
 d_3 = -d_1 \mp \frac{a^2}{1\pm d_1} = \mp 1 \pm \frac{b^2}{1\pm d_1}
  = - \frac{b^2d_1\pm a^2}{1-d_1^2}
 \ee
 so that $d_1+d_2+d_3 = \mp 1$.

 The sign ambiguity is essential for our purposes:
 {\bf only one of the two branches} (the one with $1-d_1$ in denominators)
 reproduces the right expression \cite{evo}
 for Racah matrix $\bar S_{[2]}$,
 \be
 \bar S_{[2]}
 = \frac{[2]}{[N][N+1]}
 \left(\begin{array}{ccc} 1 &  \sqrt{[N+1][N-1]}& \frac{[N]\sqrt{[N+3][N-1]}}{[2]} \\ \\
 \sqrt{[N+1][N-1]} & \frac{[N+1]}{[2][N+2]}\Big([N+3][N-1]-1\Big)
 & -\frac{[N]\sqrt{[N+3][N+1]}}{[N+2]} \\ \\
 \frac{[N]\sqrt{[N+3][N-1]}}{[2]}  & -\frac{[N]\sqrt{[N+3][N+1]}}{[N+2]}
 & \frac{[N]}{[N+2]}
 \end{array}\right) =
 \nn
 \ee
 \be
 = \frac{1}{D_1D_0}
 \left(\begin{array}{ccc}
 \frac{[2]\{q\}^2}{D_1D_0} &\frac{[2]\{q\}}{D_0}\sqrt{\frac{D_{-1}}{D_1}}
 &  \frac{\sqrt{D_3D_{-1}}}{D_1} \\ \\
 \frac{[2]\{q\}}{D_0}\sqrt{\frac{D_{-1}}{D_1}}
 & \frac{D_1}{D_2}\Big(D_3D_{-1}-\{q\}^2\Big)
 & -[2]\{q\}\frac{D_0\sqrt{D_3D_1}}{D_2} \\ \\
 \frac{\sqrt{D_3D_{-1}}}{D_1} & -[2]\{q\}\frac{D_0\sqrt{D_3D_1}}{D_2}
 & [2]\{q\}^2\frac{D_0}{D_2}
 \end{array}\right)
 \ee
 Technically this is related to factorization identity $N(N+1)-2=(N-1)(N+2)$,
 which has no analogue for $N(N+1)+2$.
 The true reason is that different branches provide matrices with two
 different signatures: $+1$ and $-1$, and only the former is the right one
 for Racah matrix $\bar S_{[2]}$.

 Thus it is not a surprise that Racah matrix $\bar S_{[1]}$ which has
 signature $0$
 is of the form (\ref{gbs2}) without any reservations:
 \be
 \bar S_{[1]}
 = \frac{1}{[N]} \left(\begin{array}{cc}
 1 & \sqrt{[N+1][N-1]}\\ \\ \sqrt{[N+1][N-1]} & -1
 \end{array}\right)
 = \frac{1}{D_0} \left(\begin{array}{cc}
 \{q\} & \sqrt{D_1D_{-1}} \\ \\ \sqrt{D_1D_{-1}} & -\{q\}
 \end{array}\right)
 \ee
 Here and further in the text we use the standard notation:
 \be
 A=q^N,\ \ \ \ \{x\} = x-\frac{1}{x}, \ \ \ \ D_k = \{Aq^k\}=Aq^k-\frac{1}{Aq^k}, \ \ \ \
 [n] = \frac{\{q^n\}}{\{q\}} = \frac{q^n-q^{-n}}{q-q^{-1}}
 \ee

 \subsection{The case of generic ${\cal N} $ }

 For higher ${\cal N}$ the out-of-diagonal constraints look like
 \be
 {\cal P}_{ij} =
 \bar s_{ij}\big(\bar s_{ii}+\bar s_{jj}\big) + \sum_{k\neq i,j}
 \bar s_{ik}\bar s_{jk} = 0 \ \ \ \ {\rm for} \ \ \ i\neq j
 \ee
 while the diagonal constraints are
 \be
 {\cal P}_{ii} = \sum_{k=1}^{\cal N} \bar s_{ik}^2 - 1 = 0
 \ee
 and it is not immediately clear which of them are actually independent.
 As we shall see in this subsection, the answer is indeed far from obvious.

 The dimension of moduli space is equal to corank of
 the $\frac{{\cal N}({\cal N}+1)}{2}\times  \frac{{\cal N}({\cal N}+1)}{2}$
 matrix  $\frac{\p {\cal P}_{ij}}{\p \bar s_{kl}}$, i.e. to the number of its
 vanishing eigenvalues -- at a point where all ${\cal P}_{ij}=0$.
 One can easily measure these eigenvalues at symmetric
 representations $R=[r]$, where the symmetric orthogonal matrix
 $\bar S$ of signature  $\ {\rm parity}(r+1)={\rm parity}({\cal N}) \ $
 is explicitly known from \cite{symreps}.
 The eigenvalues are $\pm 2$ and $0$ with the multiplicities
 \be
 \begin{array}{c|ccccccccccccc}
 r & 1 & 2 & 3 & 4 & 5 & 6 &7&8 & 9& \ldots  & r \\
 {\cal N} & 2 & 3 & 4 & 5 & 6 & 7 &8&9  &10 && r+1 \\
 \frac{{\cal N}({\cal N}+1)}{2} &  3 & 6 & 10 & 15 & 21 & 28 & 36 &45 &55&&
 \frac{(r+1)(r+2)}{2}\\
 & \\
 \#(2) & 1&3&3&6&6&10&10&15&15&&
 \frac{1}{2}\cdot \left({\rm entier}\left[\frac{r}{2}\right]+1\right)\cdot
  \left({\rm entier}\left[\frac{r}{2}\right]+2\right)\\
  &\\
 \#(-2) & 1&1&3&3&6&6&10&10&15&&
 \frac{1}{2}\cdot \left({\rm entier}\left[\frac{r-1}{2}\right]+1\right)\cdot
  \left({\rm entier}\left[\frac{r-1}{2}\right]+2\right)\\
 & \\
 \#(0) & 1&2&4&6&9&12&16&20&25&&
 r+{\rm entier}\left[\frac{r-1}{2}\right]\cdot {\rm entier}\left[\frac{r}{2}\right]
 \end{array}
 \ee
 The answer can be different for non-symmetric representations:
for $R=[2,2]$ and $R=[3,3]$ the matrices $\bar S$ have sizes $6\times 6$ and
$10\times 10$, while the eigenvalue multiplicities are
$(\#_2,\#_{-2},\#_0)=   (10,3,8)$ and   $(\#_2,\#_{-2},\#_0)=  (21,10,24)$
respectively, i.e. different from those for $R=[5]$ and $R=[9]$ with the same
sizes of $\bar S$. However, different are also the signatures:
for  $R=[2,2]$ and $R=[3,3]$ they are equal to $2$
(and further raise to $3$ for $R=[4,4]$ -- presumably it is equal to
$\ {\rm entier}\!\left[\frac{r+2}{2}\right]\ $ for $R=[r,r]$).

 However, in the case of the simplest of non-rectangular
 representations $R=[r,1]$
 the signature is just the same ${\rm parity}({\cal N})$
 as for $R=[r]$ in the case.
 Remarkably, in support of our above-presented arguments,
 the same as for $R=[9]$ is the answer for the eigenvalue multiplicity
 at the $10\times 10$
 Racah matrix $\bar S$ in representation $R=[2,1]$.

 \subsection{Conjecture about the moduli of symmetric orthogonal matrices}

 This gives certain support to the following conjecture:
 the dimension of moduli space for ${\cal N}\times{\cal N}$
 symmetric orthogonal matrices with the signature $\ {\rm parity}({\cal N})\ $ is
 \be
 {\cal D}_{\cal N} = {\cal N}-1 +
 {\rm entier}\left[\frac{{\cal N}-2}{2}\right]\cdot
 {\rm entier}\left[\frac{{\cal N}-1}{2}\right]
 \label{dimmod}
 \ee
 i.e. for large ${\cal N}$ about a quarter of the elements of $\bar S$
 are not fixed by orthogonality constraints -- twice less than for
 the ordinary orthogonal matrices.
 Still this freedom is quite big.
 It means that we should know at least
 $\frac{{\cal N}({\cal N}+1)}{2}-{\cal D}_{\cal N}$
 elements of the matrix $\bar S$ to have a chance of restoring the rest
 from orthogonality constraints, as suggested in \cite{de31}.

Of course, there is no immediate way to solve a
set of quadratic equations (unless the advanced methods of
non-linear algebra \cite{nla} are used, requiring the explicit
knowledge of the relevant resultants).
The knowledge of a part of the matrix allows to considerably
simplify this problem -- as explained in \cite{de31} it
actually  reduces  to a system of {\it linear} equations for $R=[2,1]$.
In the next section we comment on bigger representations --
there things are not so simple.
Still, we get through to the final answer at least in the case of $R=[3,1]$.

\section{Racah matrix from \cite{de31}}

Discovered in \cite{de31} was the shape of the differential expansion
\cite{IMMMfe,evo,de,db} for  colored HOMFLY-PT polynomials
of the antiparallel-double-braid knots (a certain 2-parametric generalization
of twist knots) in representations $R=[3,1]$.
After this structure is revealed, one knows the polynomials themselves
and from them one can easily read a piece  of Racah matrix $\bar S$.
It is actually entire $\bar S$ for the multiplicity-free
rectangular representations $R=[r^s]$,
but for the non-rectangular ones, beginning from $R=[r,1]$,
this is indeed a piece, moreover, a relatively small one.
Namely. extracted is a $3r \times 3r$ sub-matrix $\bar{\cal S}$ of $\bar S_{[r,1]}$,
which has the size $(7r-4)\times(7r-4)$ -- and comparison with
(\ref{dimmod}) shows that this is far below the need:
$\frac{3r(3r+1)}{2}\sim \frac{9}{2}r^2$ is parametrically much less
than the half of $\frac{(7r-4)(7r-3)}{2}\sim \frac{49}{2}r^2$.
To cure this problem it was suggested in \cite{de31} to make an
educated guess and look for $\bar S$ in the special form,
consistent with the empirical properties of the embedding
$\bar{\cal S} \embed \bar S$:

\be
\!\!\!\!\!\!\!\!\!\!\!\!\!\!\!\!\!\!\!\!\!\!\!\!\!\!
\bar S =\left(
\begin{array}{c||c|cc|ccc|cccc}
& & \tilde j = \!\!\!\!\!\!\!\!\!\!\!\!\!\!\!&\!\!\!\!\!\!\!\!\!\!\!\!\!\!\! 2,\ldots,r
& j=1,\ldots,r-1 & &&& j=\!\!\!&1,\ldots,r-1 &\\
&&&&&&&&&&\\
& 0 & \tilde j,1 & \tilde j,2  & j1 & r1 & r & j,1 & j,2 & j,3 & j,4 \\
&&&&&&&&&&\\
\hline \hline
&&&&&&&&&&\\
0 & \BS_{00} & \frac{1}{2}\BS_{0\tilde j} & \frac{1}{2}\BS_{0\tilde j}
& \BS_{0,j1} & \BS_{0,r1} & \BS_{0r} & \frac{1}{2}\BS_{0j}
& \frac{1}{2}\BS_{0j} & 0 & 0 \\
&&&&&&&&&&\\
\hline
&&&&&&&&&&\\
\tilde i,1 & \frac{1}{2}\BS_{0,\tilde i}
&  \frac{1}{4}\BS_{\tilde i\tilde j} &  \frac{1}{4}\BS_{\tilde i\tilde j}
&  \frac{1}{2}\BS_{\tilde i,j1} & 0 & \frac{1}{2}\BS_{\tilde i,r}
&\frac{1}{2}\BS_{\tilde i\,j}-x_{ij} & x_{ij}
& \frac{Y_{ij}+y_{ij}}{2} & \frac{Y_{ij}-y_{ij}}{2}
\\
&&&&&&&&&&\\
\tilde i,2 & \frac{1}{2}\BS_{0,\tilde i}
&  \frac{1}{4}\BS_{\tilde i\tilde j} &  \frac{1}{4}\BS_{\tilde i\tilde j}
&  \frac{1}{2}\BS_{\tilde i,j1} & 0 & \frac{1}{2}\BS_{\tilde i,r}
&\frac{1}{2}\BS_{\tilde i\,j}-x_{ij} & x_{ij}
& \frac{-Y_{ij}+y_{ij}}{2} & \frac{-Y_{ij}-y_{ij}}{2}
\\
&&&&&&&&&&\\
\hline
&&&&&&&&&&\\
i1 & \BS_{0,i1} & \frac{1}{2}\BS_{i1,\tilde j} & \frac{1}{2}\BS_{i1,\tilde j}
& \BS_{i1,j1} & \BS_{i1,r1} & \BS_{i1,r}
& \BS_{i1,j}-u_{ij} & u_{ij} & v_{ij} & -v_{ij}
\\
&&&&&&&&&&\\
r1 & \BS_{0,r1} & 0 & 0 & \BS_{i1,r1} & \BS_{r1,r1} & \BS_{r1,r} &
\BS_{r1,j} & 0 & 0 & 0
\\
&&&&&&&&&&\\
r & \BS_{0r} & \frac{1}{2}\BS_{r,\tilde j} & \frac{1}{2}\BS_{r,\tilde j}
& \BS_{r,j1} & \BS_{r,r1} & \BS_{r,r}
& \BS_{rj}-U_{j} & U_{j} & V_{j} & -V_{j}
\\
&&&&&&&&&&\\
\hline
&&&&&&&&&&\\
i,1 & \frac{1}{2}\BS_{0i} & \frac{1}{2}\BS_{i,\tilde j}-x_{ji}
& \frac{1}{2}\BS_{i,\tilde j}-x_{ji}
& \BS_{i,j1}-u_{ji} & \BS_{i,r1} & \BS_{ir}-U_i
& z_{ij|11} & z_{ij|12} & z_{ij|13} & z_{ij|14}
\\
&&&&&&&&&&\\
i,2 & \frac{1}{2}\BS_{0i} & x_{ji} & x_{ji} & u_{ji} & 0 & U_i
& z_{ij|21} & z_{ij|22} & z_{ij|23} & z_{ij|24}
\\
&&&&&&&&&&\\
i,3 & 0 & \frac{Y_{ij}+y_{ij}}{2} & \frac{-Y_{ij}+y_{ij}}{2} & v_{ji} & 0 & V_i
& z_{ij|31} & z_{ij|32} & z_{ij|33} & z_{ij|34}
\\
&&&&&&&&&&\\
i,4 & 0 & \frac{Y_{ij}-y_{ij}}{2} & \frac{-Y_{ij}-y_{ij}}{2} & -v_{ji} & 0 & -V_i
& z_{ij|41} & z_{ij|42} & z_{ij|43} & z_{ij|44}
\\
&&&&&&&&&&\\
\end{array}
\right)
\nn
\ee

\bigskip

This is a significant improvement: undetermined now are
the $3(r-1)^2$ parameters $x,y,Y$, $2r(r-1)$ parameters $u,v$ and
$2(r-1)(4r-3)$ parameters $z$, i.e. a total of $(13r-9)(r-1)$,
what only slightly exceeds  ${\cal D}_{7r-4}\sim \frac{49}{4}r^2$.
Moreover, some $\frac{r(r-1)}{2}$ combinations of $z$ are also
expressible through ${\cal S}$ -- thus, if all orthogonality
constraints were independent for this ansatz they would be enough.

For small $r$ such estimates are even more optimistic.

For $r=1$ there is no freedom left -- and indeed, this is a simple
case, $R=[1,1]$ is equivalent to $R=[2]$, it is sufficient to
switch $q\longrightarrow -q^{-1}$ in all the formulas.

For $r=2$ ($R=[2,1]$) there are $17$ or even $16$ free parameters (the sum of all
$z$ is a known element of $\bar{\cal S}$), what is considerably smaller
than ${\cal D}_{10} = 25$ -- and indeed,
as explained in \cite{de31}, in this case orthogonality
constraints are sufficient to restore the matrix $\bar S$ from
${\cal S}$ and the above ansatz.
This reproduces rather easily the result of \cite{GJ},
obtained by a complicated first-principle calculation.
In fact, with minor additional guesses,
coming from the desire to have factorization into quantum numbers,
in this case it is sufficient to solve only {\it linear} equations,
what makes the calculation really simple.

For $r=3$ the number of free parameters is $60$ or $57$,
while the conjecture (\ref{dimmod}) gives ${\cal D}_{17}=72$ --
thus there are also chances for success.
This time calculations require to use essentially quadratic orthogonality
equations and is pretty tedious.
In result we obtained a one-parametric family of symmetric
orthogonal matrices, with the modulus, parameterized by the angle $\theta$,
which enters only the four $Y$-parameters.
Thus, orthogonality constraints are {\it not} sufficient in this case,
even with a restrictive ansatz and with certain factorization guesses.
This angle, however, can be fixed from additional requirement --
that the eigenvalues of $\bar T^{-1}\bar S\bar T^{-1}$ are given by $T$.
Moreover, after the angle is found in this way, factorization of
the matrix elements $\bar S$ significantly improves -- what means that
the right value {\it could} be also guessed from factorization studies,
if more effort was made.

See Appendix for the list of parameters in $\bar S_{[31]}$,
which remained undetermined in \cite{de31}.

\section{Torus test and new knot polynomials}

Available test of these formulas is provided by evaluation of the
2-strand torus knots, which can be represented as 2-bridge knots,
expressible only through $\bar S$ and $\bar T$ matrices:
$${\rm Torus}(2,2k+1) =
\overline{{\rm Finger}} \Big(\underbrace{2,\ldots,2}_{2k\ {\rm times}}\Big)$$
Pictorially this looks as

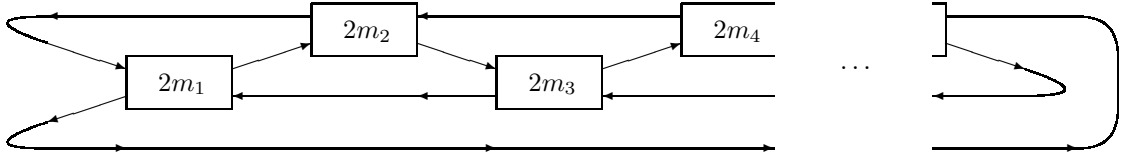
\begin{picture}(100,86)(-20,-67)
\put(30,0){\vector(-1,0){30}}
\put(0,-10){\vector(3,-1){30}}
\put(30,-30){\vector(-3,-1){30}}
\put(0,-50){\vector(1,0){30}}
\put(30,0){
\put(0,0){\line(1,0){40}}
\put(0,-50){\line(1,0){40}}
}
\put(30,-35){
\put(0,0){\line(1,0){40}}
\put(0,0){\line(0,1){20}}
\put(0,20){\line(1,0){40}}
\put(40,0){\line(0,1){20}}
\put(12,7){\mbox{$2m_1$}}
}
\put(70,0){
\put(0,0){\line(1,0){30}}
\put(0,-50){\line(1,0){30}}
}
\put(70,0){
\put(30,0){\line(-1,0){30}}
\put(0,-20){\vector(3,1){30}}
\put(70,-30){\vector(-1,0){70}}
\put(0,-50){\line(1,0){30}}
}
\put(100,-15){
\put(0,0){\line(1,0){40}}
\put(0,0){\line(0,1){20}}
\put(0,20){\line(1,0){40}}
\put(40,0){\line(0,1){20}}
\put(12,7){\mbox{$2m_2$}}
}
\put(140,0){
\put(30,0){\vector(-1,0){30}}
\put(0,-10){\vector(3,-1){30}}
\put(30,-30){\vector(-1,0){30}}
\put(-40,-50){\vector(1,0){70}}
\put(30,0){
\put(0,0){\line(1,0){40}}
\put(0,-50){\line(1,0){40}}
}
\put(30,-35){
\put(0,0){\line(1,0){40}}
\put(0,0){\line(0,1){20}}
\put(0,20){\line(1,0){40}}
\put(40,0){\line(0,1){20}}
\put(12,7){\mbox{$2m_3$}}
}
\put(70,0){
\put(0,0){\line(1,0){30}}
\put(0,-50){\line(1,0){40}}
}
\put(70,0){
\put(10,0){\line(-1,0){30}}
\put(0,-20){\vector(3,1){30}}
\put(65,-30){\vector(-1,0){65}}
\put(0,-50){\vector(1,0){65}}
}
\put(100,-15){
\put(0,0){\line(1,0){35}}
\put(0,0){\line(0,1){20}}
\put(0,20){\line(1,0){35}}
\put(12,7){\mbox{$2m_4$}}
}
}
\put(280,0){
}
\qbezier(0,0)(-30,0)(0,-10)
\qbezier(0,-50)(-30,-50)(0,-40)
\put(-40,0){
\put(340,-20){\mbox{$\ldots$}}
\put(380,0){
\put(50,0){\line(-1,0){50}}
\put(0,-10){\vector(3,-1){30}}
\put(30,-30){\vector(-1,0){35}}
\put(-5,-50){\vector(1,0){55}}
\qbezier(30,-30)(60,-30)(30,-20)
\qbezier(50,0)(65,0)(65,-15)
\qbezier(50,-50)(65,-50)(65,-35)
\put(65,-35){\line(0,1){20}}
\put(0,5){\line(0,-1){20}}
\put(0,5){\line(-1,0){5}}
\put(0,-15){\line(-1,0){5}}
}
}
\end{picture}

\noindent
where boxes contain $2m$ twists of the two lines:

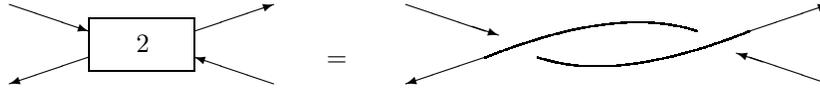
\begin{picture}(100,60)(-80,-10)
\put(0,5){
\put(0,0){\line(1,0){40}}
\put(0,0){\line(0,1){20}}
\put(0,20){\line(1,0){40}}
\put(40,0){\line(0,1){20}}
\put(18,7){\mbox{$2$}}
}
\put(-30,30){\vector(3,-1){30}}
\put(40,20){\vector(3,1){30}}
\put(0,10){\vector(-3,-1){30}}
\put(70,0){\vector(-3,1){30}}
\put(90,7){\mbox{$=$}}
\put(150,0){
\put(-30,30){\vector(3,-1){35}}
\put(100,20){\vector(3,1){30}}
\put(0,10){\vector(-3,-1){30}}
\put(130,0){\vector(-3,1){35}}
\qbezier(20,10)(50,0)(100,20)
\qbezier(0,10)(50,30)(80,20)
}
\end{picture}

\noindent
Expression for HOMFLY-PT polynomial,
obtained by the standard rules of \cite{arbor}, is
\be
H_R^{\overline{{\rm Finger}}_R(2m_1,2m_2,\ldots)} =
d_R\cdot \Big(\bar S \bar T^{2m_1} \bar S \bar T^{2m_2} \bar S \ldots
\bar T^{2m_{2k}}\bar S\Big)_{\emptyset\emptyset}
\ee
Bar over "finger" reminds that it involves only crossings of anti-parallel lines.
Since knot polynomials for torus knots are known for arbitrary representation
$R$ from the Rosso-Jones formula \cite{RJ,DMMSS},
one can make a comparison -- and it is indeed successful.
After that one can immediately calculate $[31]$-colored polynomials
for all knots with arbitrary parameters $m_1,m_2,\ldots$ --
they are all 2-bridge and thus not too many.
It looks plausible,
but is not quite unclear if {\bf  all the 2-bridge can be brought to this form},
with antiparallel crossings only.
Still this produces quite a few new results.
The simplest single-antiparallel-finger knots are:

$$\!\!\!\!\!\!\!\!\!\!\!\!\!\!\!\!\!\!\!\!\!\!\!\!\!\!\!\!\!\!
\begin{array}{c|c||c|c||c|c||c|c||c|c}
  {\rm knot} & \{m\} & {\rm knot} & \{m\} & {\rm knot} & \{m\} & {\rm knot}
& \{m\} & {\rm knot} & \{m\}\\
\hline
&&&&&&&&&\\
3_1 & \underline{\underline{\boxed{1,1}}}   &
8_1 & \underline{\underline{1,-3}} &
9_1 & \boxed{1,1,1,1,1,1,1,1} &  10_1 & \underline{\underline{1,-4}} & 10_{25} & ?
\vspace{-0.1cm}
\\
&&
8_2 & 1,1,1,1,1,-1 &
9_2 & \underline{\underline{1,4}} & 10_2 & 1,1,1,1,1,1,1,-1 & 10_{26} & ?
\\
4_1 & \underline{\underline{1,-1}}&
8_3 & \underline{2,-2} &
9_3 & 2,1,1,1,1 & 10_3 & \underline{2,-3} &10_{27}& ?
\\
&&
8_4 & 2,-1,-1,-1 &
9_4 & 1,1,1,3 & 10_4 & 1,1,1,-3 &10_{28}& ?
\\
5_1 &  \boxed{1,1,1,1} &
8_6 & 1,1,2,-1 &
9_5 & \underline{3,2} &  10_5 & 1,1,1,1,1,1,-1,1 &10_{29}& ?
\\
5_2 & \underline{1,2}  &
8_7 & 1,1,1,1,-1,-1 &
9_6 & 1,1,1,1,2,1 &  10_6 & 1,1,1,1,2,-1 &10_{30}& ?
\\
&&
8_8&1,1,-2,-1 &
9_7 & 1,1,3,1 & 10_7 & 3,1,1,-1 &10_{31}&?
\\
6_1 & \underline{\underline{1,-2}} &
8_9 &1,1,1,-1,-1,-1&
9_8 & 1,1,-2,1 &  10_8 & 1,1,1,1,1,-2 &10_{32}& ?
\\
6_2 & 1,1,1,-1 &
8_{11} & ? &
9_9 & 1,1,1,2,1,-1 & 10_9 & 1,1,1,1,1,-1,-1,-1 &10_{33}& ?
\\
6_3 & 1,1,-1,-1 &
8_{12} & 1,-1,1,-1 &
9_{10} & ? & 10_{10} & 3,1,-1,-1 &10_{34}&?
\\
&&
8_{13} & ? &
9_{11} & 1,-1,1,1,1,1 &  10_{11} & 1,1,2,-2 &10_{35}& 1,-1,2,-1
\\
7_1 & \boxed{1,1,1,1,1,1} &
8_{14} & ? &
9_{12} & 2,-1,1,1 &  10_{12} & 1,1,1,2,1,1 &10_{36} &?
\\
7_2 & \underline{\underline{1,3}} &&&
9_{13} & 2,2,1,1 & 10_{13} & 1,-1,1,-2  & 10_{37} &?
\\
7_3 & 1,1,1,2 & 8_5 & &
9_{14} & ? & 10_{14} & 1,1,1,2,1,-1 &10_{38} & ?
\\
7_4 & \underline{2,2} &8_{10}&3-{\rm bridge}&
9_{15} & 1,-1,2,1 & 10_{15} &1,1,1,1,-2,-1 &10_{39}&?
\\
7_5 & 1,1,2,1 &8_{15}-&{\rm knots}& 9_{17} & ? & 10_{16} & ? & 10_{40} & ?
\\
7_6 & 1,1,-1,1 &-8_{21}&& 9_{18} & 2,1,2,1 & 10_{17} & ? & 10_{41} &  ?
\\
7_7 & 1,-1,-1,1 & && 9_{19} & 1,-1,-2,1 & 10_{18} & ? &10_{42} & ?
\\
&&&&9_{20}&?&10_{19}&? &10_{43}& ? \\
&&&&9_{21}&?&10_{20}& 1,1,3,-1 &10_{44} & ? \\
&&&&9_{23}& ?  &10_{21} &? &10_{45} & ?   \\
&&&&9_{26}& ? &10_{22}& ? &&   \\
&&&&9_{27}& ? &10_{23}& ? &10_{46}-& 3-{\rm bridge} \\
&&&&9_{31}& ? &10_{24}& ? &-10_{165}& {\rm knots}
\end{array}
$$

\newpage
\noindent
Every knot in the table has many realizations of this kind,
we include only the simplest one.
Underlined are the knots, describable by
only two non-vanishing parameters $m_1$ and $m_2$ -- these
are {\it double braids}, which possess a remarkable factorization
of differential-expansion coefficients into those for twist knots
(double-unerlined), and were the source of knowledge about the
sub-matrix $\bar{\cal S}$ from \cite{de31}.
Sensitive to the other elements of $\bar S$
(though not to the angle $\theta$, see the Appendix) are
non-underlined knots.
For testing our formulas the torus knots were used -- the simplest of
them  are present in the table and marked by boxes.
Omitted are the knots which are not 2-bridge, i.e. not representable
as single fingers with both types of crossings allowed, parallel and antiparallel.
The ones which are not yet(?) identified as single {\it antiparallel} fingers
are labeled by question marks.

The next immediate things to do are extraction of the matrix $S$ from
(\ref{SthrbS}) and development of arborescent calculus {\it a la}
\cite{arbor} for representation $R=[3,1]$.


\section*{Appendix}

We list here the entries of $\bar S$ for representation $R=[3,1]$,
complementing the matrix elements of the non-orthogonal sub-matrix ${\cal S}$,
which was derived in \cite{de31} from the newly-discovered
differential expansion of the $[r,1]$-colored HOMFLY-PT polynomials.

\be
\begin{array}{l|l}
x_{21} = \frac{\{q\}^2}{D_0^2}\cdot\sqrt{\frac{D_{-2}}{D_2}}
\cdot\Big(Aq^2+\frac{1}{Aq^2}\Big), \ \ \ \ \
&x_{22} = -\frac{D_3D_1-\{q\}^2}{[2]D_0}\cdot\sqrt{\frac{D_{-2}}{D_3D_2D_1}}\nn \\
\nn \\
y_{21} = \frac{[2]\{q\} }{D_0^2 \sqrt{[2]D_3D_2}}\cdot
\Big(2D_3D_{-1}-D_0^2+\{q\}^2\Big) \ \ \ \
&  y_{22} = -\frac{1}{[2]D_0}\cdot\sqrt{\frac{D_4}{D_3D_2D_1}}\cdot
\Big(D_1^2-D_1D_{-1}-2[3]\{q\}^2\Big) \ \ \ \ \nn \\ \nn\\
Y_{21} = \cos\theta = \{q\}\sqrt{\frac{[2]}{D_3D_2}}\ \ \ \
&Y_{22} = \sin\theta = \sqrt{\frac{D_4D_1}{D_3D_2}}
\nn\\   \\
x_{31}=-\frac{1}{[3]D_0}\cdot \sqrt{\frac{[2]D_4D_1D_{-2}}{D_2}}, \ \ \ \ \
&x_{32} = \frac{\{q\}}{[3]D_1}\cdot \sqrt{\frac{D_4D_{-2}}{[2]D_3D_2}}
\nn \\
y_{31} = \frac{2D_2-[3]D_0}{[3]D_0}\cdot\sqrt{\frac{D_4D_1}{D_3D_2}} \ \ \ \
& y_{32} = - \frac{\{q\}}{[3]D_1 \sqrt{[2]D_3D_2}}\cdot
\left([4]D_1 - \{q\}\Big(Aq+\frac{1}{Aq}\Big)\right)
\nn \\ \nn \\
Y_{31} = -\sin\theta = -\sqrt{\frac{D_4D_1}{D_3D_2}} \ \ \
&Y_{32} = \cos\theta = \{q\}\sqrt{\frac{[2]}{D_3D_2}}
\end{array}
\ee
Matrix $\bar S_{[3,1]}$ is symmetric and orthogonal for arbitrary value of parameter
$\theta$. Moreover, $\theta$ does not contribute to expressions
for 2-strand torus knots and other single-finger knots, considered in sec.5.
However, it affects the eigenvalues of $\bar T^{-1}\bar S{\bar T}^{-1}$
and is fixed by comparison with the entries of $T$.
\be
\begin{array}{l|l}
u_{11} = \frac{\{q\}D_3}{D_2D_0}\cdot \sqrt{\frac{D_{-3}}{D_{-1}}} \ \ \ \ \
& v_{11} =\frac{\{q\}}{D_2D_0}\cdot \sqrt{\frac{[2]D_3D_{-1}D_{-3}}{D_{-2}}} \nn \\
u_{12} = \frac{D_3}{[2]D_2}\cdot \sqrt{\frac{D_3D_{-3}}{D_1D_{-1}}} \ \ \ \ \
& v_{12} = -\frac{1}{[2]D_2}\cdot\sqrt{\frac{D_4D_3D_{-1}D_{-3}}{D_1D_{-2}}} \nn \\
\nn \\
u_{21} = \frac{D_4}{[3]D_2D_0^2}\cdot\sqrt{D_3D_1D_{-1}D_{-3}} \ \ \ \ \
&v_{21} = \frac{D_4D_{-1}}{[3]D_2D_0^2}\cdot
\sqrt{\frac{[2]D_1D_{-1}D_{-3}}{D_{-2}}}\nn \\
u_{22} = -\frac{\{q\}D_4}{[3][2]D_2D_1D_0}\cdot\sqrt{D_{-1}D_{-3}} \ \ \ \ \
&v_{22} =\frac{\{q\}\big([4]D_1+D_0\big)}{[3][2]D_2D_1D_0}\cdot
\sqrt{\frac{D_4D_{-1}D_{-3}}{D_{-2}}} \nn \\
\nn \\
U_1 = \frac{[2]D_{-2}}{[3]D_2D_0}\cdot\sqrt{D_5D_1} \ \ \ \ \
&V_1 = -\frac{1}{[3]D_0}\cdot\sqrt{\frac{[2]D_5D_1D_{-2}}{D_3}}\nn \\
U_2 = -\frac{\{q\}D_{-2}}{[3]D_2D_1}\cdot\sqrt{\frac{D_5}{D_3}} \ \ \ \ \
&V_2 = -\frac{\{q\}}{[3]D_1} \cdot  \sqrt{ \frac{D_5D_{-2}}{D_4D_3} }
\end{array}
\ee
The $z$-constituents of $\bar S$ are listed in an order, which reflects
their hidden symmetry:
\be
\bar S_{10,10} &= \frac{[2]\{q\}^2}{D_3D_2D_0^2D_{-2}}\cdot
\Big(
[2]D_4D_0D_{-3}+D_2D_0^2 - [2]\{q\}^2(2D_1+D_{-1})
\Big) \nn \\
\bar S_{10,11} &= \frac{[2]\{q\}^2}{D_2D_0^2}\cdot\Big(D_3+[2]D_{-2}\Big)\nn \\
\bar S_{10,14} &= \frac{\{q\}}{D_2D_0D_{-2}\sqrt{D_3D_1}}\cdot
\Big(D_5D_0D_{-3}+D_4D_0D_{-2}-2[2]\{q\}^2D_1\Big)\nn \\
\bar S_{10,15} &= \frac{\{q\}}{D_2D_0 \sqrt{D_3D_1}}\cdot
\Big(\frac{[4]}{[2]}D_4D_0 + \{q\}^2\Big)
\nn
\ee

$$
\begin{array}{l|lcl}
\bar S_{11,11} = \frac{\{q\}^2}{D_2D_0^2}\cdot\Big([2]D_3-D_0\Big)
& \bar S_{11,14} = \frac{\{q\}\big(D_4+[2]^2D_{-2}\big)}{[2]D_2D_0}
\cdot\sqrt{\frac{D_3}{D_1}}
&& \bar S_{11,15} = -\frac{\{q\}\big([2]D_1+D_0\big)}{[2]D_2D_0}
\cdot\sqrt{\frac{D_3}{D_1}}   \\
& &\ \ \ &\\
&\bar S_{12,14} = -\frac{\{q\}\big([2]D_{-1}-D_2\big)}{D_2D_0}\cdot
\sqrt{\frac{D_1}{[2]D_{-2}}}
&& \bar S_{12,15} = - \frac{\{q\}}{D_2D_0}\cdot\sqrt{\frac{D_1D_{-2}}{[2]}} \\
&&&\\
&\bar S_{13,14} = -\bar S_{12,14}
&& \bar S_{13,15} = -\bar S_{12,15} \\
&&&
\end{array}
$$
$$
\bar S_{14,14} = D_5D_2D_0^2-[2]\{q\}^2\Big([4][3]D_5D_0 + ([3]+\{q\}^2)D_4D_0\Big)
-[4]^2[2]\{q\}^4  + [2]^3\{q\}^6
$$
$$
\bar S_{14,15} = \frac{D_5D_{-2} - \frac{[6]}{[3]}\{q\}^2}{[2]^2D_2D_1}
 \ \ \ \ \ \ \ \ \ \ \ \  \ \ \ \   \ \ \ \ \ \ \ \   \ \ \ \ \ \ \ \ \ \ \ \
\bar S_{15,15} = \frac{D_1D_0+[2]\{q\}^2}{[2]^2D_2D_1} \ \ \ \ \ \ \ \ \
\nn \\
$$

$$
\!\!\!\!\!\!\!\!\!\!\!\!\!\!\!\!\!
\begin{array}{ll|ll}
&& \\
\bar S_{10,12} = \frac{[2]^2\{q\}^3}{D_2D_0^2\sqrt{[2]D_3D_{-2}}}
\Big(qA^2-\frac{1}{qA^2}\Big)
&\bar S_{10,13} = -\bar S_{10,12}
&\ \ \bar S_{10,16} = -\frac{\{q\}^2}{D_2D_0}\sqrt{\frac{D_4}{D_3D_1D_{-2}}}
\Big(qA^2-\frac{1}{qA^2}\Big)
&\bar S_{10,17} = -\bar S_{10,16}\nn \\
&&\\
\bar S_{11,12} = \frac{ \{q\}^2}{D_2D_0^2\sqrt{[2]D_3D_{-2}}}
&\bar S_{11,13} = -\bar S_{11,12}
&\bar S_{11,16} = -\frac{\{q\}}{[2]D_2D_0}\sqrt{\frac{D_4D_3D_{-2}}{D_1 }}
&\bar S_{11,17} = -\bar S_{11,16}\nn \\
&&\\
\bar S_{12,12} = -\frac{ [2]\{q\}^2}{D_2D_0^2}
&\bar S_{12,13} = -\bar S_{12,12}
&\bar S_{12,16} = -\frac{\{q\}}{[2]D_2D_0}\sqrt{[2]D_4 D_1 }
&\bar S_{12,17} = -\bar S_{12,16}\nn \\
&&\\
\bar S_{13,12} = - \bar S_{12,12}
&\bar S_{13,13} =  \bar S_{12,12}
&\bar S_{13,16} = -\bar S_{12,16}
&\bar S_{13,17} = \bar S_{12,16}\nn \\
&&\\
&& \bar S_{14,16} = \frac{D_5D_0^2-[2]\{q\}^4D_2}{[2]^2D_2D_1\sqrt{D_4D_{-2}}}
&\bar S_{14,17} =  -\bar S_{14,16} \\
&&\\
&&\bar S_{15,16} =  \frac{ \sqrt{D_4 D_{-2}} }{[2]^2D_2}
&\bar S_{15,17} = -\bar S_{12,16}\nn \\
&&\\
&& \bar S_{16,16} =  \frac{D_2D_1+[2]\{q\}^2}{[2]^2D_2D_1}
&\bar S_{16,17} =  -\bar S_{16,16} \\
&&\\
&&\bar S_{17,16} = - \bar S_{16,16}
&\bar S_{17,17} = \bar S_{16,16}\nn \\
\end{array}
$$

\section*{Acknowledgements}

I am indebted for discussions and essential help to
A.Mironov, An.Morozov, Sh.Shakirov and A.Sleptsov.

\noindent
This work was funded by the Russian Science Foundation (Grant No.16-12-10344).

\end{document}